\documentclass[aps,twocolumn,prd,showpacs,nofootinbib]{revtex4}
\usepackage{amsmath}
\usepackage{graphicx}
\usepackage{dcolumn}
\usepackage{bm}
\usepackage{amssymb}
\usepackage{latexsym}

\def\setC{\mathbb{C}}

\def\setN{\mathbb{N}}

\def\setR{\mathbb{R}}

\newcommand{\dd}{\mathrm{d}}

\newcommand{\ie}{\textsl{i.e.~}}
\newcommand{\cf}{\textsl{cf.~}}
\newcommand{\eg}{\textsl{e.g.~}}
\newcommand{\etal}{\textit {et al.~}}
\newcommand{\apriori}{\textit{a priori~}}

\newcommand{\GN}{G_{_{\rm N}}}
\newcommand{\mP}{M_{_{\mathrm P}}}
\newcommand{\GReCO}{${\cal G}\setR\varepsilon\setC{\cal O}$}

\newcommand{\rdim}{r}
\newcommand{\radim}{\rho}
\newcommand{\F}{{\cal F}}

\begin{document}

\title{Dilatonic current-carrying cosmic strings}

\author{Patrick Peter}
\email{peter@iap.fr}
\affiliation{Institut d'Astrophysique de
Paris, \GReCO, FRE 2435-CNRS, 98bis boulevard Arago, 75014 Paris,
France}

\author{M.~E.~X.~Guimar\~aes\footnote{On leave from Universidade de
Bras\'{\i}lia, Departamento de Matem\'atica}}
\email{emilia@ift.unesp.br}
\affiliation{Instituto de F\'{\i}sica Te\'orica, Universidade Estadual
Paulista, Rua Pamplona 145, 01405-900 S\~ao Paulo SP, Brazil}

\author{V. C. de Andrade}
\email{andrade@ift.unesp.br}
\affiliation{Instituto de F\'{\i}sica Te\'orica, Universidade Estadual
Paulista, Rua Pamplona 145, 01405-900 S\~ao Paulo SP, Brazil}

\date{May, 23$^\mathrm{rd}$ 2003}

\begin{abstract}
We investigate the nature of ordinary cosmic vortices in some
scalar-tensor extensions of gravity. We find solutions for which the
dilaton field condenses inside the vortex core. These solutions can be
interpreted as raising the degeneracy between the eigenvalues of the
effective stress-energy tensor, namely the energy per unit length $U$
and the tension $T$, by picking a privileged spacelike or timelike
coordinate direction; in the latter case, a {\sl phase frequency
threshold} occurs that is similar to what is found in ordinary neutral
current-carrying cosmic strings. We find that the dilaton
contribution for the equation of state, once averaged along the string
worldsheet, vanishes, leading to an effective Nambu-Goto behavior of
such a string network in cosmology, \ie on very large scales. It is
found also that on small scales, the energy per unit length and
tension depend on the string internal coordinates in such a way as to
permit the existence of centrifugally supported equilibrium
configuration, also known as {\sl vortons}, whose stability, depending
on the very short distance (unknown) physics, can lead to catastrophic
consequences on the evolution of the Universe.
\end{abstract}


\maketitle

\section{Introduction}

Most extensions of the standard model of particle physics predict that
extra scalar fields, in addition to the ordinary Higgs field, whose
experimental detection is still to be done, should exist in Nature. At
low energies (compared to the Planck scale), they appear to be
classifiable into essentially two main categories, namely those which
couple in a straightforward way to the other particle fields (as, \eg
grand unification breaking Higgs fields, supersymmetric partners of
ordinary fermions or extra bosonic degrees of freedom coming from the
Neveu-Schwartz sector of superstring theory~\cite{sugrastring}), and
those whose most important coupling is to gravity, such as the
dilaton, whose origin can be traced to the Ramond sector in the
superstring context. Both kinds, coupled or decoupled, have been
studied from different (and often disjoint) perspectives, and both
have various cosmological and astrophysically observable consequences;
these terms permit us, for instance, to obtain fully nonsingular
cosmologies~\cite{FFPP}. In particular, scalar-tensor theories of
gravity~\cite{ST} may provide a natural solution to the problem of
terminating inflation~\cite{endinflation}, whilst grand unified theory
(GUT) scalars, being symmetry breakers, may lead to the formation of
topological defects~\cite{TD}, of which only cosmic strings are viable
candidates from the point of view of cosmology.

Among these theories, some predict both kinds of fields. As a result,
one expects that cosmic strings could exist whose coupling to gravity
would be altered by inclusion of dilaton effects. In
Ref.~\cite{emilia}, a local cosmic string solution was considered in
the framework of low energy string theory which is reminiscent of
the scalar-tensor theories of gravity~\cite{ST}.  Indeed, a massless
dilaton is shown to obey a least coupling principle~\cite{DP}, \eg
to decouple from matter by cosmological attraction in much the same
way as the generic attractor mechanism of the scalar-tensor theories
of gravity~\cite{DN}. It was found~\cite{emilia} then that the metric
around a cosmic string in the framework of scalar-tensor gravity is of
the Taub-Kasner type~\cite{TK} so that the particle and light
propagation resembles that around a wiggly cosmic string in ordinary
general relativity~\cite{wiggly}, although the effect was expected to
be one order of magnitude stronger. 

Here, we want to point out another effect, namely that the dilaton
field may behave as a winding phase along the string, thereby
generating a neutral current kind of effect by raising the degeneracy
between the eigenvalues of the stress-energy tensor. {}From the point
of view of purely gravitational physics, this seems utterly negligible
as the metric would hardly be affected~\cite{UnotT,fer} by such a
current (it gives again a Kasner-like metric, up to second order
corrections).

The most noticeable consequence of a current-like effect
is~\cite{witten,formal,neutral} to modify the internal dynamics of
cosmic strings in such a way that new states are reachable. Indeed,
the breaking of the Lorentz boost invariance along the worldsheet
allows rotating equilibrium configurations, called vortons, which, if
they are stable, can overclose the universe, thereby leading to a
catastrophe for the theory that predicts them~\cite{vortons}. Finally,
inclusion of such an internal structure could drastically change the
predictions of a cosmic string model~\cite{rdp} in the microwave
background anisotropies~\cite{CMB}. Here we show that the long-range
effect on a cosmologically relevant network of strings is vanishing
on average, but that vorton-like states can be reached by
microscopically small loops.

In what follows, after having set the relevant gravitational theory
and notations, we derive the corresponding field equations in
Sec.~\ref{Sec:frame}. We then set the vortex (Abelian Higgs) model
which we develop in flat space in Sec.~\ref{Sec:vortex} with the aim
of using it as a source for the gravitational effects. We then move on
to obtain, in this framework, the general solution for the dilaton
field in the Einstein frame (Sec.~\ref{Sec:dila}). We apply this
solution to derive the effective stress-energy tensor
(Sec.~\ref{Sec:effT}) of the string, as seen from a Jordan-Fierz frame
observer. We show that this stress-energy tensor has very particular
features that can be interpreted by saying that a network of such
string will evolve on cosmologically relevant scales as a usual
network~\cite{network} of Nambu-Goto strings~\cite{GN}, but might lead
to the formation of equilibrium {\sl vorton} states~\cite{vortons}
whose density, scaling as matter, could overclose the Universe in the
case in which they are stable, an issue which is yet unresolved,
depending on the small distance physics. Section \ref{Sec:conc}
summarizes our findings and discusses the relevant cosmological
conclusions.

\section{Gravitational framework} \label{Sec:frame}

We start with the gravitation action in the Jordan-Fierz frame (also
known as the ``string'' frame, a nomenclature we shall not use in
order to avoid the possible risk of confusion with the strings of the
cosmic kind we consider below), namely
\begin{eqnarray}
{\cal S}_{_{\mathrm JF}} &=& \frac{1}{16\pi} \int \dd^4 x
\sqrt{-\tilde{g}} \left[ \tilde{R} \tilde{\Phi} -
\frac{\omega(\tilde{\Phi})}{\tilde{\Phi}}\partial^{\mu}\tilde{\Phi}
\partial_{\mu}\tilde{\Phi}\right] \nonumber \\ & & + {\cal S}_{\mathrm
m}\left[\Psi_{\mathrm m} (x), \tilde{g} _{\mu\nu} (x)\right], \label{JF}
\end{eqnarray}
where $\tilde{g}_{\mu\nu}$ is the physical metric which contains
both scalar and tensor degrees of freedom, $\tilde{R}$ is the
curvature scalar associated with it, and ${\cal S}_{\mathrm m}$ is
the action for general matter fields $\Psi_{\mathrm m}$ which, at
this point, is left arbitrary. The metric signature is assumed to
be $(+,-,-,-)$.

By varying the action (\ref{JF}) with respect to the metric
$\tilde{g}_{\mu\nu}$ and to the scalar field $\tilde{\Phi}$ we obtain
the ``modified" Einstein equations, and a wave equation for
$\tilde{\Phi}$, namely
\begin{eqnarray}
\tilde{G}_{\mu\nu} & = & \frac{8\pi }{\tilde{\Phi}}\tilde{T}_{\mu\nu}
+ \frac{1}{\tilde{\Phi}}\left(\tilde\nabla_{\nu}\tilde{\Phi}_{,\mu} -
\tilde{g} _{\mu\nu}\stackrel{\sim}{\Box}\tilde{\Phi} \right)
\nonumber \\ & & + \frac{\omega(\tilde{\Phi})}{\tilde{\Phi}
^2}\left(\partial_{\mu}\tilde{\Phi} \partial_{\nu}\tilde{\Phi}
-\frac{1}{2}
\tilde{g}_{\mu\nu}\partial_\alpha\tilde{\Phi}\partial^\alpha\tilde{\Phi}
\right),\label{eqEin} \\ \stackrel{\sim}{\Box}\tilde{\Phi} & = &
\frac{1}{2\omega(\tilde{\Phi}) + 3} \left( 8\pi \tilde{T} -
\frac{\dd\omega}{\dd\tilde{\Phi}}\partial_{\mu}\tilde{\Phi}
\partial^{\mu}\tilde{\Phi} \right)\!\!, \label{eqmot} \\
\tilde\nabla_{\mu}\tilde{T}^{\mu}_{\ \ \nu} & = & 0 \label{dt0}
\end{eqnarray}
where a tilde over a differential operator means it is built out of
the Jordan-Fierz metric $\tilde g_{\mu\nu}$,
\begin{equation}
\tilde{G}_{\mu\nu} = \tilde{R}_{\mu\nu} -
\frac{1}{2}\tilde{g}_{\mu\nu}\tilde{R}\label{G}
\end{equation}
is accordingly the Einstein tensor in the Jordan-Fierz frame, and
\begin{equation}
\tilde{T}_{\mu\nu} \equiv \frac{2}{\sqrt{-\tilde{g}}} \frac{\delta
{\cal S}_{\mathrm m}}{\delta \tilde{g}^{\mu\nu}} \label{tmunu}
\end{equation}
is the energy-momentum tensor of the matter content and $\tilde{T}
\equiv \tilde{T}^{\mu}_{\ \mu}$ is its trace. Clearly, if $\tilde{T}$
vanishes and $\tilde{\Phi}$ is a constant, Eq.~(\ref{eqEin}) reduces
to the usual set of Einstein field equations if we identify the
inverse of the scalar field with the Newton constant, \ie $\GN =
1/\tilde{\Phi}$. Hence, any exact solution of Einstein equations with
a trace-free matter source will also be a particular exact solution of
the scalar-field with $\tilde{\Phi}$ constant. Of course, this
particular solution will not be, except in very special situations,
the general solution for the matter content~\cite{baw}.

Let us rewrite the action given by Eq.~(\ref{JF}) in terms of the
Einstein (conformal) frame in which the kinematic terms of tensor and
scalar degrees of freedom do not mix, \ie
\begin{eqnarray}
{\cal S}_{_{\mathrm E}} &=& \frac{1}{16\pi G^*} \int \dd^4x
\sqrt{-g} \left( R - 2g^{\mu\nu} \partial_{\mu}\phi
\partial_{\nu}\phi \right) \nonumber \\ & & + {\cal S}_{\mathrm m}
\left[\Psi_{\mathrm m}, A^2(\phi)g_{\mu\nu}\right] , \label{EF}
\end{eqnarray}
where $g_{\mu\nu}$ is a pure rank-2 metric tensor, $R$ is the
curvature scalar associated to it, and $G^*$ the bare gravitational
constant.

As is well known, the action given by Eq.~(\ref{EF}) is obtained from
that of Eq.~(\ref{JF}) by means of a conformal transformation
\begin{equation}
\tilde{g}_{\mu\nu} = A^2(\phi) g_{\mu\nu}, \label{conftrans}
\end{equation}
provided the scalar field functions $\phi$ and $\tilde \Phi$ are
related through
\begin{equation}
G^* A^2(\phi) = \frac{1}{\tilde{\Phi}},\label{phiPhi}
\end{equation}
and
\begin{equation}
\frac{\dd \ln A(\phi)}{\dd\phi} =
\frac{1}{\sqrt{2\omega(\tilde{\Phi}) + 3}}  \equiv \alpha(\phi),
\end{equation}
(thus defining the function $\alpha$) which can be interpreted as the
(field-dependent) coupling strength between matter and scalar field.

In the conformal frame, Eqs. (\ref{eqEin}) and ({\ref{eqmot}) are
written in a more convenient form
\begin{equation}
G_{\mu\nu}= 2\partial_{\mu}\phi\partial_{\nu}\phi -
g_{\mu\nu}g^{\alpha\beta}\partial_{\alpha}\phi\partial_{\beta}\phi
 + 8\pi G^* T_{\mu\nu},\label{Einstein} 
\end{equation}
for the gravitational part, and
\begin{equation}
\Box\phi = - 4\pi G^* \alpha(\phi) T,
\label{dilaton}
\end{equation}
for the dilatonic part, where now the matter stress-energy tensor
$T_{\mu\nu}$ is obtained from
\begin{equation}
T_{\mu\nu} \equiv \frac{2}{\sqrt{-g}}\frac{\delta {\cal
S}_{\mathrm m}}{\delta g^{\mu\nu}}, \label{EFtmunu}
\end{equation}
which in this new frame is no longer conserved unless the dilaton is
constant, \ie $\nabla_{\mu} T^{\mu}_{\ \ \nu} = \alpha(\phi)
T\nabla_{\nu}\phi$. The Einstein frame Einstein tensor $G_{\mu\nu}$
appearing in Eq.~(\ref{Einstein}) is defined in the same was as
Eq.~(\ref{G}) without the tildes. {}From Eq.~(\ref{conftrans}), we can
easily relate quantities from both frames in the following way:
\begin{equation}
\tilde{T}_{\mu\nu} = A^{-2}(\phi) T_{\mu\nu}, \label{JFtoE}
\end{equation}
which also implies $\tilde{T}^\mu_{\ \nu} = A^{-4} T^\mu_{\ \nu}$
and $\tilde{T}^{\mu\nu} = A^{-6} T^{\mu\nu}$. For the sake of
generality, we choose to leave $A(\phi)$ as an arbitrary function
of the scalar field.

Let us now turn to the cosmic string source terms and consider the
microscopic field theory out of which vortices stem.

\section{Vortex field model} \label{Sec:vortex}

We shall now consider the underlying field model that gives birth
to cosmic strings. It consists in a complex scalar Higgs field
$\varphi$, coupled to a gauge vector $B_\mu$. Both fields are, as
discussed above, minimally coupled to gravity so that the matter
action we shall deal with is expressible as

\begin{equation}
{\cal S}_{\mathrm m} = \int \dd^4 x \sqrt{-\tilde g} \left(
{\frac{1}{2}}|D \varphi|^2 - {\frac{1}{4}} H_{\mu\nu} H^{\mu\nu}
- V \right), \label{U1model}
\end{equation}
where the U(1) covariant derivative is $D_\mu \equiv \partial_\mu
+ iq B_\mu$, the ``Faraday''-like tensor $H_{\mu\nu}\equiv
\partial_\mu B_\nu -
\partial_\nu B_\mu$, and the Higgs potential reads $V(\varphi ) =
\lambda (\varphi^\star \varphi - \eta^2)^2$; all indices are
raised and lowered by means of the metric $\tilde g$.

We shall from now on consider the zeroth order approximation for the
background fields. This means we are interested in the string as a
source for the gravitational and dilaton fields. As a result, in order
to derive the relevant stress-energy tensor, we demand that the
Einstein-frame metric be that of Minkowski, while the dilaton assumes
a constant value, \ie
\begin{equation}
\tilde\Phi_{(0)} = \tilde\Phi_0 \Longrightarrow \phi_{(0)} = \phi_0,
\end{equation}
so that, at this order, gravity is described by general relativity in
both frames and the Jordan-Fierz metric can be taken, in a cylindrical
coordinate system $(t,z,\rdim,\theta)$, as
\begin{equation}
\tilde g^{\mu\nu}_{(0)} = A_0^{-2} \eta^{\mu\nu} = A_0^{-2} \
\hbox{Diag } \left (1,-1,-1,-\frac{1}{\rdim^2}\right),
\label{metric0}
\end{equation}
\ie again the Minkowski metric [up to a constant scaling factor $A_0
\equiv A(\phi_0)$]. The usual Newton constant is then $\GN =
\tilde\Phi_0^{-1} = G^* A_0^2$. Note that in Eq.~(\ref{metric0}) we
have inserted the (constant) conformal factor in the Jordan-Fierz
metric: this is just for further convenience since we will be mostly
working in the Einstein frame in which this extra factor will then be
absent.

There exist static vortex configurations that are solutions of the
Euler-Lagrange equations derivable from the action given by
Eq.~(\ref{U1model}). Such a configuration, for a string along the $z$
axis, has the form~\cite{nielsen}
\begin{equation}
\varphi = h (\rdim) \hbox{e}^{i n \theta}, \ \ \ \hbox{and} \ \ \ B_\mu
= {\frac{1}{q}} \left[ Q(\rdim ) -n\right] \delta^\theta_\mu,
\label{confstring}
\end{equation}
where the functions $h$ and $Q$ depend on the radial distance to
the string core $\rdim$ only. In what follows, for the sake of
definiteness, we shall also assume that the underlying parameters
in the matter action are such that only the vortices with winding
number $n=1$ are stable and we shall therefore concentrate our
attention on these configurations. Note however that this
requirement will not modify our conclusions, since what is
presented here is merely an existence proof that only relies on
the presence of the defect itself.

Using a prime to denote differentiation with respect to the radial
distance $\rdim$, the field equations derivable from the
action~(\ref{U1model}) are
\begin{equation}
h^{\prime\prime}+ {\frac{h^{\prime}}{\rdim}} = h \left[
{\frac{Q^2}{\rdim^2}} + 4\lambda A_0^2 (h^2 - \eta^2)\right]
\label{Higgs}
\end{equation}
and
\begin{equation}
Q^{\prime\prime}- {\frac{Q^{\prime}}{\rdim}} = q^2 Q A_0^2 h^2,
\label{gauge}
\end{equation}
and the boundary conditions for these fields to describe a vortex line
read
\begin{equation}
\left\{
\begin{array}{ll}
h(0) = 0, & Q(0) = 1, \\ & \\ \displaystyle{\lim_{\rdim\to\infty}} h
(\rdim) =\eta, & \displaystyle{\lim_{\rdim\to\infty}} Q(\rdim) =0.
\end{array}
\right.  \label{boundary}
\end{equation}
The field equations~(\ref{Higgs}) and (\ref{gauge}), together
with the conditions (\ref{boundary}) are usually solved
numerically; an example of such a solution is shown on
Fig.~\ref{fit}, adapted from Ref~.~\cite{neutral}. On the figure
are shown the dimensionless quantities
\begin{equation}
X\equiv \frac{h}{\eta} \label{Xdef}
\end{equation}
and $Q(\radim)$ as functions of the rescaled (dimensionless)
distance to the string core
\begin{equation}
\radim\equiv\frac{\rdim}{\rdim_{\mathrm h}},\ \ \ \ \ \hbox{ where } \
\ \ \ \rdim_{\mathrm h} \equiv \lambda^{-1/2} \eta^{-1}
\label{Comptondef}
\end{equation}
is the Compton wavelength of the Higgs field. Such generic
configurations are the source for the dilaton field.

In order to derive the internal string solution for the dilaton field
in which we are interested, we need to first obtain the stress-energy
tensor (\ref{EFtmunu}), namely
\begin{eqnarray}
\tilde T_{\mu\nu} &=& \frac{1}{2} \left[\left( D_{\mu}
\varphi\right)^\dagger D_{\nu}\varphi + \left( D_{\nu}
\varphi\right)^\dagger D_{\mu}\varphi \right] - \frac{1}{2} \tilde
g_{\mu\nu} |D\varphi |^2 \nonumber \\ & & + \tilde g_{\mu\nu}
V(\varphi) +\frac{1}{4} \tilde g_{\mu\nu} H^2 - \tilde
g^{\alpha\beta} H_{\mu\alpha} H_{\nu\beta},\label{Ttildemunu}
\end{eqnarray}
and, setting $V\equiv V(h) = \lambda \left( h^2 -\eta^2 \right)^2$,
this yields the following components
\begin{eqnarray}
\tilde T^z_{\ z} &=& V+ {\frac{A_0^{-2}}{2}} \left[ {h^{\prime}}^{2}
+{\frac{1}{\rdim^2}} \left( h^2 Q^2 + {\frac{{Q^{\prime}}^2}{q^2
A_0^2}} \right) \right],
\label{tttzz} \\
\tilde T^\rdim_{\ \rdim} &=& V - {\frac{A_0^{-2}}{2}} \left[
{h^{\prime}}^2 -{\frac{1}{\rdim^2}} \left( h^2 Q^2 -
{\frac{{Q^{\prime}}^2}{q^2 A_0^2}}\right) \right],
\label{trhorho}\\
\tilde T^\theta_{\ \theta} &=& V + {\frac{A_0^{-2}}{2}} \left[
{h^{\prime}}^2
-{\frac{1}{\rdim^2}} \left( h^2 Q^2 + {\frac{{Q^{\prime}}^2}{q^2
A_0^2}}\right) \right],
\label{tthetatheta}
\end{eqnarray}
and $\tilde T^t_{\ t} = \tilde T^z_{\ z}$. This zeroth order
stress-energy tensor should in principle be used as a source for the
modified Einstein equations.

It can be noted that, as is clear from Eqs.~(\ref{Higgs}) and
(\ref{gauge}) as well as (\ref{tttzz}) to (\ref{tthetatheta}),
the normalization $A_0$ of the dilaton function $A(\phi)$ can be
modified at will provided one performs simultaneously a
redefinition of the coupling constants $\lambda$ and $q$ through
$\bar\lambda=A_0^2\lambda $ and $\bar q= A_0 q$. In fact, this
normalization turns out to be completely irrelevant for the
vortex configurations since all the properties of such vortices
only depend on the ratio~\cite{neutral} $\lambda/q^2 =
\bar\lambda/\bar q^2$. This stems from the fact that $A_0^2
\tilde T^\mu_{\ \nu}$ can be expressed in terms of $\bar\lambda$
and $\bar q$ only, so the only effect is a normalization one. It
is therefore possible in principle to set $A_0=1$, a convention
within which the metrics in either the Jordan-Fierz or the
Einstein frame are exactly equal. In order to distinguish between
these frame, we shall however not adopt this convention, unless
stated otherwise.

\begin{figure}[t]
\hskip-1.5cm
\includegraphics[width=10cm]{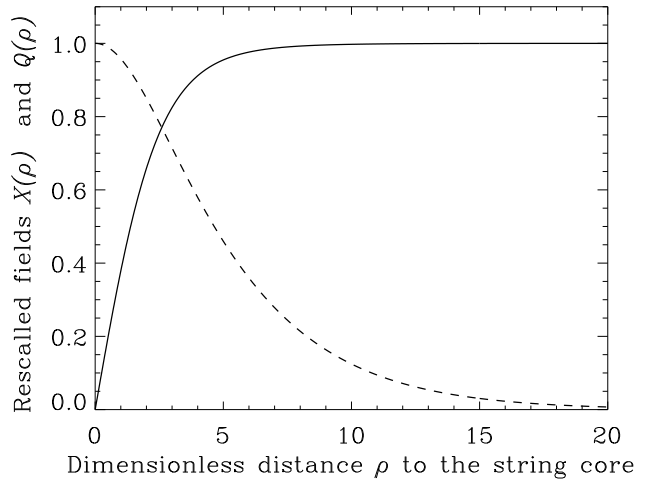}
\caption{Numerical solutions for the rescaled Higgs (full line) and
gauge (dashed line) fields around a vortex in a Minkowski background
with $A_0=1$. Adapted from Ref.~\cite{neutral}.} \label{fit}
\end{figure}

\section{First order dilaton solution}\label{Sec:dila}

We now switch to the Einstein frame. The stress-energy tensor just
derived then provides the new frame one through the relation
(\ref{JFtoE}), so that the trace needed in Eq.~(\ref{dilaton})
thus takes the form $T=A^4 \tilde T$. Inserting
Eqs.~(\ref{tttzz})--(\ref{tthetatheta}) into Eq.~(\ref{dilaton}), we
obtain the dynamical equation, up to first order in the
gravitational constant, for the dilaton as
\begin{equation}
\Box \phi = - 4\pi G^* \alpha(\phi_0) A_0^2 \left[ h'^2
+\frac{h^2 Q^2}{\rdim^2} +4\lambda A_0^2 \left( h^2 -
\eta^2\right)^2 \right], \label{phiinterne}
\end{equation}
which may be expressed as
\begin{equation}
\Box \phi = \epsilon \F (r), \label{boxphi_in}
\end{equation}
where we have set $\alpha_0\equiv \alpha(\phi_0)$. The function $\F$
on the right hand side of Eq.~(\ref{boxphi_in}) is given, in terms of
the dimensionless quantities, by [see Eq.~(\ref{Xdef}) and below]
\begin{equation} 
\F = - \frac{\alpha_0 A_0^2}{\rdim_{\mathrm h}^2} \left[ \left(
\frac{\dd X}{\dd \radim} \right)^2 + \frac{X^2 Q^2}{\radim^2} + 4
A_0^2 \left( X^2-1 \right)^2\right], \label{calF}
\end{equation}
and is exhibited in Fig.~\ref{sourcephi}.

In Eq.~(\ref{boxphi_in}), we have emphasized the constant
combination
\begin{equation}
\epsilon \equiv 4 \pi G^* \eta^2 \sim 4 \pi \left( \frac{\eta}
{\mP}\right)^2, \label{epsilon}
\end{equation}
which will be used in what follows as a small expansion
parameter. Indeed, even for the highest possible energy phase
transition leading to cosmic strings compatible with cosmological
data~\cite{bprs}, \ie the GUT scale, the quantity $\eta$ is of order
$10^{15}-10^{16}$ GeV, which is at most three orders of magnitude
smaller than the Planck scale $\mP\equiv \GN^{-1/2}$ so that one has
$\epsilon \alt 10^{-5}$.

\begin{figure}[t]
\hskip-1.5cm
\includegraphics[width=10cm]{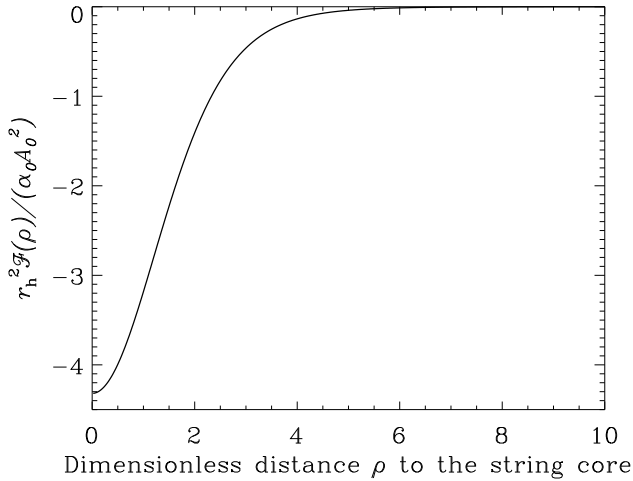}
\caption{Source function for the dilaton in Eq.~(\ref{boxphi_in})
with the vortex solution of Fig~.~\ref{fit}.} \label{sourcephi}
\end{figure}

Let us now expand all the fields involved in terms of the small
parameter $\epsilon$. In what follows, we shall concentrate on the
dilaton field, because the solution of the Einstein equations
(\ref{Einstein}) for the metric has already been obtained in the
cosmic string case~\cite{emilia}. In this reference, it had been found
that the external metric, far from the string core, was of the
Kasner-like form
\begin{equation}
\dd s^2 = \left(\frac{\rdim}{\rdim_0} \right)^k \left( \dd t^2 -
\dd z^2\right) - \dd\rdim^2 - \Gamma \rdim^2
\left(\frac{\rdim}{\rdim_0} \right)^{-2k} \dd\theta^2,
\label{metric1}
\end{equation}
with $\Gamma$ and $k$ two arbitrary constants, and that it could be
matched with the interior solution provided $k=4/3$ or $k=0$ in the
presence of the string.

{}From Fig.~\ref{sourcephi}, it is clear that there exists a distance
$\rdim_0$ such that for $\rdim \geq \rdim_0$, the source function is
approximately vanishing, so that the exterior solution for the dilaton
should satisfy
\begin{equation}
\Box \phi_\mathrm{ext} =0. \label{boxphi_out}
\end{equation}
Therefore, the dilaton field, in vacuum ($T_{\mu\nu}=0$), assumes the
general form~\cite{emilia}
\begin{equation}
\phi_\mathrm{ext} = \phi_0 + \kappa \ln
\left(\frac{\rdim}{\rdim_0}\right), \label{phi0}
\end{equation}
where $\phi_0$ and $\kappa$ are constant, the latter being determined
by a matching with the internal solution, while the former gives the
strength of gravitational coupling.

In the absence of a string, one would have $\kappa=0$, as demanded
also by the requirement that the Ricci tensor be
regular~\cite{emilia}. This solution, as it turns out~\cite{emilia},
is valid, to first order in the weak field approximation, both inside
and outside the string. We will thus use this solution to derive the
string structure itself, and show, for self-consistency, that a
modification of the dilaton solution with respect to
Ref.~\cite{emilia} does not modify this metric (again, at least to
first order in $\epsilon$). The mild (logarithmic) divergence observed
in Eq.~(\ref{phi0}) for the dilaton far from the string stems from the
infinite string approximation we are making use of, and can easily be
accounted for by introducing a long range cutoff such as, \eg the
curvature radius of the string, or the interstring distance in a
cosmological network. We shall see later that it can also be
altogether cancelled once the source term is taken into account [see
the discussion below Eq.~(\ref{nolog})].

It is interesting to note here that the solution (\ref{phi0}) for the
dilaton actually also diverges as $\rdim\to 0$, implying a breakdown
of the underlying four-dimensional effective field theory. This is
similar to the situation encountered when the axion field is taken in
consideration in cosmic strings formed at the symmetry breaking of the
pseudo-anomalous U(1) that characterizes~\cite{binetruy} most cases of
superstring compactification~\cite{pseudo}, indicating that
topological defect cores might be objects of comparable theoretical
interest as black hole or cosmological singularities in that they
probably require a full knowledge of the non linear theory to be
properly understood. In what follows, we shall assume a short-range
cutoff for the dilaton, expected to be of order $\mP^{-1}$, and
subsequently neglect distances shorter than the Planck length;
inclusion of this cutoff scale merely renormalizes the string energy
per unit length and tension by factors of order unity~\cite{binetruy}
that are irrelevant to the following discussion. Another implication
of this divergence is that some cosmologically interesting effects,
such as formation of wakes by dilatonic strings, may break down due to
the logarithmic divergence of this solution~\cite{oliveira}.

Assuming the dilaton to behave as
\begin{equation}
\phi = \phi_0 + \epsilon \phi_1, \label{expansion}
\end{equation}
where, as before, $\phi_0$ is the constant dilaton value in the
absence of string, and $\phi_1$ depends on the radial coordinate
$\rdim$ as well as the string coordinates $z$ and $t$ (we assume
rotational symmetry so that $\partial_\theta\phi_1
=0$), Eq.~(\ref{boxphi_in}) inside the string, up to first order in
$\epsilon$, becomes
\begin{equation}
\left(\ddot \phi_1 - \bar{\bar{\phi}}_1 \right) - {\frac{1}{\rdim}}
{\frac{\partial}{\partial\rdim}} \left(\rdim
{\frac{\partial\phi_1}{\partial\rdim}}\right) = \F (r),
\label{boxphi1}
\end{equation}
where a dot and a bar respectively stand for derivations with
respect to the coordinates $t$ and $z$.

As we want to match the solution of Eqs.~(\ref{expansion}) and
(\ref{boxphi1}) with the already derived solution (\ref{phi0}),
we seek the following form:
\begin{equation}
\phi_1 (t,\rdim,z) = \chi (\rdim) + f(\rdim) \psi(z,t) ,
\label{temptative}
\end{equation}
where the function $f(\rdim)$ is required to vanish asymptotically far
from the defect, \ie in practice for $\rdim\geq \rdim_0$, in order to
ensure that the corresponding effect is localized into the worldsheet
only. Note that the $z$ and $t$ dependence of the dilaton in
Eq.~(\ref{temptative}) is not incompatible with the assumption of
time-independence and cylindrical symmetry for the metric at the first
order in $\epsilon$. Indeed, as can readily be seen in
Eqs.~(\ref{Einstein}) and~(\ref{expansion}), the leading contribution
in the metric of the correction $\phi_1$ is second order in
$\epsilon$.  With such a tentative solution, Eq.~(\ref{boxphi1})
reduces to
\begin{equation}
\left(\ddot \psi - \bar{\bar{\psi}}\right) f -
\frac{1}{\rdim}\frac{\dd}{\dd\rdim}\left(
\rdim \frac{\dd\chi}{\dd\rdim} \right)- 
\frac{1}{\rdim}\frac{\dd}{\dd\rdim}\left(
\rdim \frac{\dd f}{\dd\rdim} \right)
\psi = \F (\rdim).
\label{boxphidev}
\end{equation}
In order for our solution to be valid regardless of the behavior of
$\psi(z,t)$, \ie including the case $\psi=0$, we demand that the
function $\F$ sources only the pure radial component of the dilaton,
\ie we impose
\begin{equation}
\chi''+{\frac{1}{\rdim}}\chi'= \F(r), \label{chieq}
\end{equation}
which implies
\begin{equation}
\chi = A + B\ln \left(\frac{\rdim}{\rdim_0}\right) +
\chi_{_\mathrm{S}},
\label{solchi}
\end{equation}
where $A$ and $B$ are two arbitrary constants and the special solution 
\begin{equation}
\chi_{_\mathrm{S}}=\int^\rdim_{\rdim_1}
\frac{\dd\tilde\rdim}{\tilde\rdim}\int^{\tilde\rdim}_{\rdim_2}\F
(\bar\rdim) \bar\rdim \dd\bar\rdim, \label{chi}
\end{equation}
depends \apriori on the two constants $\rdim_1$ and $\rdim_2$. Note
however that $\rdim_1$ and $\rdim_2$ have no physical influence and
can be chosen at will. In particular, it is convenient to set
$\rdim_1=\rdim_2=0$, so that, with $\F\sim \F_0 +\F_2 \rdim^2$, which
is the short distance behavior of $\F$ (see Fig.~\ref{sourcephi}), one
gets
\begin{equation}
\chi \sim A + B\ln \left(\frac{\rdim}{\rdim_0}\right)
+\frac{\F_0}{4}\rdim^2 +\frac{\F_2}{9}\rdim^3 +\cdots,
\end{equation}
close to the string core, \ie in the limit $\rdim\to 0$. The special
solution $\chi_{_\mathrm{S}}$ cannot alleviate the divergence of the
dilaton near the string core.

On the other hand, Eq.~(\ref{chi}) can also be used to match the
exterior solution (\ref{phi0}) to the interior solution. In
particular, it is interesting to note that for large distances, and
because the function $\F$ vanishes exponentially fast,\footnote{This
can be seen through a careful examination of the asymptotic behaviors
of the various fields involved, as derived, \eg in Ref.~
\cite{neutral}.} one has, for large values of $\rdim$, \eg $\rdim >
\rdim_\infty$ with $\rdim_\infty$ far away from the string core,
\begin{eqnarray}
\chi_{_\mathrm{S}} &=& \int^{\rdim_\infty}_{0}
\frac{\dd\tilde\rdim}{\tilde\rdim}\int^{\tilde\rdim}_{0}\F (\bar\rdim)
\bar\rdim \dd\bar\rdim+\int_{\rdim_\infty}^{\rdim}
\frac{\dd\tilde\rdim}{\tilde\rdim}\int^{\tilde\rdim}_{0}\F (\bar\rdim)
\bar\rdim \dd\bar\rdim\nonumber\\ &\simeq&\hbox{ f.p. } +
\left[\int_0^\infty \F (\tilde\rdim) \tilde\rdim \dd\tilde\rdim\right]
\ln\left(\frac{\rdim}{\rdim_\infty}\right), \label{nolog}
\end{eqnarray}
where ``f.p.'' stands for the finite part of the above
relation. Because the constant $B$ in Eq.~(\ref{solchi}) is, at this
stage, arbitrary, it can be chosen to exactly compensate for the
asymptotic logarithmic divergence in $\chi_{_\mathrm{S}}$, in such a
way that the exterior solution for the dilaton can be consistently
imposed to be a constant, \ie $\kappa=0$ in Eq.~(\ref{phi0}). As
opposed to any other choice, this one leads to a finite amount of
energy. This is reminiscent of what happens around the vortices
studied in Ref.~\cite{binetruy}, whose coupling with the axion made
them local even though they were initially global.

Now, returning to Eq.~(\ref{boxphidev}) in which we insert the
solution for $\chi$ and separating, we obtain
\begin{equation}
{\frac{1}{\psi}}\left(\ddot{\psi} - \bar{\bar{\psi}}\right)
= {\frac{1}{f}}\left(f^{\prime\prime}+{\frac{1}{\rdim}}
f^{\prime}\right) = w,
\end{equation}
with $w$ a constant. Therefore, we have the following set of
equations:
\begin{equation}
\ddot \psi - \bar{\bar{\psi}} = w \psi , \label{psi}
\end{equation}
and
\begin{equation}
f^{\prime\prime}+{\frac{1}{\rdim}} f^{\prime}= w f,
\label{f}
\end{equation}
for $f$ satisfying the boundary condition $\lim_{\rdim\to\infty}
f(\rdim) =0$. The arbitrary constant $w$ can assume \apriori both
positive and negative values, so we will inspect both cases in turn
later, but from now on let us consider Eq.~(\ref{psi}) for the phase
modulation $\psi$ depending on the variables $(t,z)$, and seek a
solution of the form
\begin{equation}
\psi(t,z) = \psi(kz-\omega t)\equiv\psi(u) ,\label{modulation}
\end{equation}
which gives
\begin{equation}
\left(\omega^2-k^2\right)\frac{\dd^2\psi}{\dd u^2} = w\psi.
\end{equation}
An overall rescaling of the coordinates being always possible, we can
without lack of generality assume that $w=\pm
\left(\omega^2-k^2\right)$. The positive sign choice however leads to
exponentially growing or decaying solutions in the variable $u$ which
are either unbounded or vanishing, hence physically irrelevant. We are
thus led to impose the negative sign, namely
\begin{equation}
w=k^2-\omega^2,\label{w}
\end{equation}
and the solution for the dilaton
\begin{equation}
\psi = \psi_\mathrm{s}\sin\left(kz-\omega t\right) +
\psi_\mathrm{c}\cos\left(kz-\omega t\right).\label{phase}
\end{equation}
It is worth noting that if $w\geq 0$ (respectively $w\leq 0$), the
variable $u$ defines a new spacelike (resp. timelike) coordinate. In
order to simplify the following calculations, we shall perform a
Lorentz boost along the string such that if $w\geq 0$ (respectively
$w\leq 0$), the new time and space coordinates $t'$ and $z'$ and the
corresponding new constants $\omega'$ and $k'$ are such that $u=k' z'$
(resp. $u=-\omega' t'$) and $w=k'^2$ (resp. $w=-\omega'^2$). Assuming
this new frame from now on, we will then drop the primes as there is no
risk of confusion.

Consider first the case for which $w = -\omega^2$, \ie a negative
constant. It can be seen that Eq.~(\ref{f}) then becomes the Bessel
equation of order zero, with general solution~\cite{Grad}
\begin{equation}
f(\rdim) = f_J J_0 (\omega \rdim) + f_Y Y_0 (\omega
\rdim), \label{time}
\end{equation}
with \apriori arbitrary numerical coefficients $f_J$ and
$f_Y$, $J_0$ and $Y_0$ being respectively Bessel functions of
the first and second kind. The boundary condition that $f$ should
vanish asymptotically is not enough to impose any condition on the
choice of the constants; even the fact that $Y_0$ diverges near the
axis does not lead to any new constraint since $Y_0(\omega\rdim)
\propto \ln (\omega\rdim)$, \ie a divergence similar to that already
observed for the radial part $\chi(\rdim)$, for which a cutoff need be
imposed at the Planck scale. However, as we shall show later, there
are other constraints stemming from the requirement that the
eigenvalues of the energy-momentum tensor, once integrated in the
directions transverse to the string, be finite.

Let now $w$ be a positive constant, \ie $w = k^2$. Equation (\ref{f})
is in this case the modified Bessel function of order zero, with
solution
\begin{equation}
f(\rdim) = f_I I_0 (k \rdim) + f_K K_0 (k
\rdim). \label{space}
\end{equation}
Here again, the constants $f_I$ and $f_K$ are \apriori
arbitrary and must be designed in such a way as to match the exterior
solution. It is clear however that since we demand the dilaton first
order correction to vanish asymptotically, we must impose
$f_I=0$ since $I_0$ is exponentially divergent for large
arguments. Note that here as well as in the previous case
(\ref{time}), the solution involves a logarithmic divergence
reminiscent of the behavior given by Eq.~(\ref{phi0}), whose
significance is discussed underneath that equation.

Equations (\ref{time}) and (\ref{space}), together with
Eq.~(\ref{phase}), appear to completely determine the space-time
behavior of the dilaton field in all possible situations. We now turn
to the consequences that this solution produces in the effective
stress-energy tensor.

\section{Effective Stress-Energy tensor} \label{Sec:effT}

We now consider the effective stress-energy tensor $\tilde
T_\mathrm{(eff)}^{\mu\nu}$ that is seen by an observer wishing to
describe the string behavior in the framework of Einstein general
relativity. This means $\tilde T_\mathrm{(eff)}^{\mu\nu}$ is given by
assuming Eq.~(\ref{eqEin}) takes the form
\begin{equation}
\tilde R_{\mu\nu} - \frac{1}{2}\tilde g_{\mu\nu}\tilde R = 8\pi \GN
\tilde T_{\mu\nu}^\mathrm{(eff)},
\label{Reff}
\end{equation}
a relation that we will use later to identify the effective energy per
unit length and tension of the string to first order in $\epsilon$. In
order to achieve this goal, let us remark that Eq.~(\ref{phiPhi})
implies that
\begin{equation}
\frac{1}{\tilde\Phi}\sim G^* A_0^2 \left( 1+2\epsilon \alpha_0
\phi_1\right),
\label{1surPhi}
\end{equation}
\ie $\tilde\Phi \propto \left( 1-2\epsilon \alpha_0 \phi_1\right)$,
which depends on space and/or time coordinates only through the first
order dilaton field $\phi_1$. Plugging this form back into
Eq.~(\ref{Reff}), keeping in mind that $\GN \tilde
T_\mathrm{(eff)}^{\mu\nu}$ is already a first order quantity and that
$\partial_\mu \tilde \Phi = -2\alpha_0 \epsilon \partial_\mu
\phi_1/\GN$, we find that, to first order, the effective stress-energy
tensor we are seeking reads
\begin{equation}
8\pi\tilde T_{\mu\nu}^\mathrm{(eff)} = 8\pi\tilde
T^\mathrm{(0)}_{\mu\nu} +\left( \partial_\mu\partial_\nu
-\Gamma^{\alpha}_{\ \mu\nu}\partial_\alpha-\eta_{\mu\nu}\Box\right)
\tilde\Phi +{\cal O}^\mathrm{(2)}_{\mu\nu}, 
\label{Tmunueff}
\end{equation}
where $\tilde T^\mathrm{(0)}_{\mu\nu}$ is the zeroth order
stress-energy tensor given by Eqs.~(\ref{tttzz})--(\ref{tthetatheta})
and ${\cal O}^\mathrm{(2)}_{\mu\nu}$ contains only terms proportional
to $\epsilon^2$.

In order to determine the influence of the dilaton field on the string
dynamics, let us first recall the relevant pieces of formalism needed
to describe it from the macroscopic point of view~\cite{formal}. We
shall consider our string to be describable by means of a surface
action and accordingly integrate the effective stress-energy tensor
over the transverse degrees of freedom, \ie
\begin{equation}
\tilde T^{\mu\nu}_{||} \equiv \int \tilde T^{\mu\nu}_\mathrm{(eff)} \,
\dd^2x^{\perp},
\label{Tpara}
\end{equation}
where $\dd^2x^{\perp}$ accounts for the transverse measure around the
string. To the required zeroth order (since the integrand $\tilde
T_{\mu\nu}^\mathrm{(eff)}$ is of first order) and given the symmetry
in the solution, this is $\dd^2x^{\perp} = 2\pi \rdim\, \dd
\rdim$. The macroscopic stress-energy tensor $\tilde {\bf
T}_{_\mathrm{M}}$, depending only on the internal string coordinates
$\xi_a$, is derivable from the relation
\begin{equation}
\tilde T^{\mu\nu}_{||}(x^\alpha) = \int \tilde
T^{\mu\nu}_{_\mathrm{M}} \delta\left[ x^\alpha -
X^\alpha\left(\xi_a\right)\right]\, \dd^2\xi,
\label{macroT}
\end{equation}
with $a=0,1$ and $X^\alpha\left(\xi_a\right)$ defining the
two-parameter locus of the string. In the case at hand for which the
string is aligned along the $z$ axis, the coordinates $\xi_a$ can be
identified with $t$ and $z$; we shall make this choice in what
follows.

Since one expects $\tilde T_\mathrm{(eff)}^{\mu\nu}$ to be
conserved\footnote{Note in that respect that Eq.~(\ref{eqmot}) can be
seen as a simple consequence of Eqs.~(\ref{eqEin}) and (\ref{dt0}),
\ie of the conservation of the matter stress-energy tensor and that of
Einstein tensor.} by virtue of Eq.~(\ref{eqEin}), reproducing the
steps of Ref.~\cite{UnotT} leads to the fact that $\tilde {\bf
T}_{_\mathrm{M}}$ can only depend on two integrated quantities. It
turns out that it can in fact be given the form
\begin{equation}
\tilde T_{_\mathrm{M}} = U u^\mu u^\nu - T v^\mu v^\nu,
\label{TM}
\end{equation}
with ${\bf u}$ and ${\bf v}$ respectively a unit timelike and
spacelike vector parallel to the string worldsheet; again, in our
case, these are ${\bf u}= (1,0,0,0)$ and ${\bf v}= (0,1,0,0)$ [recall
Eq.~(\ref{metric0})].

The eigenvalues of $\tilde {\bf T}_{_\mathrm{M}}$ are the energy per
unit length $U$ and the tension $T$, which we are now in a position to
express directly from the effective stress-energy tensor as
\begin{equation}
U = 2\pi \int \tilde T_\mathrm{(eff)}^{tt} \rdim\,\dd \rdim,
\label{U}
\end{equation}
and 
\begin{equation}
T = -2\pi \int \tilde T_\mathrm{(eff)}^{zz} \rdim\,\dd \rdim.
\label{Tension}
\end{equation}
Turning back to Eq.~(\ref{Tmunueff}), it is straightforward to convince
oneself that these quantities take the form
\begin{equation}
U = m_0^2 + 2\pi \int \partial_t^2 \tilde \Phi \, \rdim\,\dd \rdim, \
\ \ T = m_0^2 - 2\pi \int \partial_z^2 \tilde \Phi \, \rdim\,\dd
\rdim,
\label{UTPhi}
\end{equation}
where $m_0^2$ is an integral over the transverse direction of the
part of the microscopic fields that depends only on the radial distance
$\rdim$, \ie a constant with the dimension of a mass square (hence the
notation).

Using the solutions (\ref{temptative}) and (\ref{phase}) together with
the expansion (\ref{1surPhi}), we get
\begin{equation}
U = m_0^2 + 4\pi \alpha_0 \omega^2 \epsilon \psi(kz-\omega t)\int
f(\rdim) \, \rdim\,\dd \rdim,
\label{U1}
\end{equation}
and
\begin{equation}
T = m_0^2 - 4\pi \alpha_0 k^2 \epsilon \psi(kz-\omega t)\int
f(\rdim) \, \rdim\,\dd \rdim,
\label{T1}
\end{equation}
relations that imply that not only is the stress-energy tensor no
longer degenerate when inclusion of the dilatonic field is taken into
account, but also that the resulting string is not of the Witten
superconducting kind~\cite{witten,neutral} since the energy per unit
length and tension explicitly depend on the internal coordinates.

{}From Eqs.~(\ref{U1}) and (\ref{T1}) and the solution (\ref{time}),
one sees immediately that because of the asymptotic behavior of the
Bessel functions $J_0$ and $Y_0$, the timelike case $w<0$ is excluded
for the case at hand since it leads to divergent integrals in $U$ and
$T$. Indeed, the integrals \{see Eq.~(5.52/1) in Ref.~\cite{Grad}\}
are\footnote{We call ${\cal Z}_p$ any Bessel function, modified or
not, of order $p$.} proportional to $\int \rdim {\cal Z}_0 \left(
\omega\rdim\right) \dd \rdim = (\rdim/\omega) {\cal Z}_1 \left(
\omega\rdim\right)$, which asymptotically behaves as
$\sqrt{\rdim_\infty}$, with $\rdim_\infty$ an appropriate cutoff, for
the timelike case~(\ref{time}), and exponentially converges for the
spacelike case~(\ref{space}).

The timelike case can however be accounted for by adding a potential
term for the dilaton, \ie by substituting the Einstein frame action
(\ref{EF}) with
\begin{eqnarray}
{\cal S}^\mathrm{ (pot)}_{_{\mathrm E}} &=& \frac{1}{16\pi G^*} \int
\dd^4x \sqrt{-g} \left[ R - 2g^{\mu\nu} \partial_{\mu}\phi
\partial_{\nu}\phi + V\left(\phi\right)\right]\nonumber \\ & & +
{\cal S}_{\mathrm m} \left[\Psi_{\mathrm m},
A^2(\phi)g_{\mu\nu}\right] , \label{EFpot}
\end{eqnarray}
which turns Eq.~(\ref{dilaton}) into
\begin{equation}
\Box\phi = -\frac{1}{4}\frac{\dd V}{\dd \phi} - 4\pi G^* \alpha(\phi) T.
\label{dilatonpot}
\end{equation}
Defining the dilaton mass $M_\mathrm{d}$ through 
\begin{equation}
M_\mathrm{d}^2 \equiv \frac{1}{4}\frac{\dd^2 V}{\dd
\phi^2}\Bigg|_{\phi=\phi_0},
\end{equation}
and following the same steps as in Sec.~\ref{Sec:dila}, we find that
the first order dilaton $\phi_1$ satisfies
\begin{equation}
\left(\ddot \phi_1 - \bar{\bar{\phi}}_1 \right) - {\frac{1}{\rdim}}
{\frac{\partial}{\partial\rdim}} \left(\rdim
{\frac{\partial\phi_1}{\partial\rdim}}\right) + M_\mathrm{d}^2 \phi_1
= \F (r),
\label{boxphi1pot}
\end{equation}
instead of Eq.~(\ref{boxphi1}). To obtain Eq.~(\ref{boxphi1pot}), we
have made use of the fact that in this context, $\phi_0$ is the vacuum
expectation value of $\phi$ and the dilaton is stabilized in the sense
that $\dd V/\dd\phi =0$ for $\phi=\phi_0$ and that
$M_\mathrm{d}^2>0$. Making use again of the expansion
(\ref{temptative}) and choosing the purely radial part to satisfy
\begin{equation}
\chi''+\frac{1}{\rdim}\chi'-M_\mathrm{d}^2 \chi = \F(r), \label{chieqpot}
\end{equation}
instead of Eq.~(\ref{chieq}), one sees that Eq.~(\ref{psi}) is left
unchanged, implying the same solution (\ref{phase}), while
Eq.~(\ref{f}) is turned into
\begin{equation}
f''+\frac{1}{\rdim} f'= \left( M_\mathrm{d}^2 + w \right) f,
\label{fpot}
\end{equation}
whose solution is an exponentially convergent modified Bessel function
[\cf Eq.~(\ref{phase})] provided the phase frequency, in the frame in
which $k\to 0$, is below the threshold~\cite{neutral} $\omega<
\omega_\mathrm{th} = M_\mathrm{d}$. It is remarkable that this
threshold is, just like in the Witten superconducting string
model~\cite{witten}, also set by the mass of what one could thus, by
analogy, call the {\sl current carrier}.

{}From the point of view of cosmology, the stress-energy tensor
eigenvalues given by Eqs.~(\ref{U1}) and (\ref{T1}) depend on the
internal worldsheet coordinates, but in a very special way. In fact,
defining the average of the quantity $X$ over the spacelike or
timelike variable $u=kz-\omega t$ by means of
\begin{equation}
\langle X \rangle \equiv \frac{1}{2\pi}\int_{-\pi}^{\pi} X(u)\dd u,
\end{equation}
we obtain the very simple Nambu-Goto~\cite{GN} relation
\begin{equation}
\langle U \rangle = \langle T \rangle = m_0^2,
\end{equation}
showing that on distances much larger than the characteristic dilaton
length scale, for instance over distances of cosmological relevance, a
network of such cosmic strings will evolve in a way that is similar to
ordinary (non-current-carrying) strings. In particular, one therefore
expects the overall network to rapidly reach a scaling solution and to
produce a large number of small loops~\cite{network}. At this level,
all the cosmological predictions of the ordinary string models are
unchanged.

Once we consider the smaller loops, however, the situation can be
drastically modified, in a fashion comparable to the vorton
case~\cite{vortons}: since the degeneracy between $U$ and $T$ is
raised microscopically, one might also expect centrifugally supported
equilibrium configurations to exist, leading to the usual vorton
excess problem. Indeed, the presence of an explicit phase factor
$\psi$ in their definition implies that the quantization condition in
the spacelike case
\begin{equation}
k L = 2\pi N,
\end{equation}
with $N\in\setN$ and $L$ the loop circumference, must hold. It is not
clear however if the dilaton can leave the string, and the mechanism
by which it could be possible presumably depends on what happens below
the cutoff at short distance.

Finally, one can note that the difference between the eigenvalues $U$
and $T$ is $U-T\propto \psi$, which, given Eq.~(\ref{phase}), is not
positive definite. This can be traced back to the well-known expected
violation of the null energy condition (NEC) in such scalar-tensor
theories. A similar violation allows cosmological solutions to have
bouncing scale factors with ordinary matter components (fluids and/or
scalar fields)~\cite{FFPP,stringbounce}. In the string case under
consideration here, this seemingly acausal violation of the NEC stems
from the coupling of the dilaton field to the matter fields. The
causality issue here stems from the fact that in the usual treatment
of string perturbations~\cite{formal}, the velocity of transverse
perturbations along the string is found to be $c_{_\mathrm{T}}^2
\equiv T/U$, which clearly exceeds unity if the NEC is
violated. However bizarre it may sound, this need not worry us
unduly. Indeed, since the theory given by Eq.~(\ref{JF}) satisfies the
requirement of causality however, it is clear that $c_{_\mathrm{T}}^2$
cannot, in this framework, represent the propagation velocity of
transverse perturbations along the string.

\section{Conclusions} \label{Sec:conc}

We have reexamined the field equations of a cosmic string coupled to a
tensor-scalar theory of gravity. This coupling is shown to induce
effects along the string comparable to a current flow in the sense
that the resulting effective stress-energy tensor eigenvalues, the
energy per unit length $U$ and tension $T$, are no longer degenerate,
due to the presence of the dilaton. However, we have found that there
are many differences with the ordinary mechanism of current formation
in cosmic strings as was first proposed by Witten~\cite{witten}. In
the latter case, it was shown~\cite{neutral} that the generated
currents can be either of the spacelike kind or of the timelike
kind. In the dilatonic situation under scrutiny here, the timelike
case is found to be pathological and can be accounted for only by
addition of a potential term for the dilaton. Once this is done
however, one finds that the phase frequency threshold derived in the
Witten case~\cite{neutral} has an exact counterpart since the timelike
kind of ``current'' can only be reached provided the equivalent of the
state parameter is larger than the dilaton mass.

Another difference concerns the fact that the current is not formed
after the string forming phase transition but should instead appear
exactly at the same time. This is due to the fact that the dilaton is
not an ordinary scalar field but acts as a component of the
gravitational interaction. From this peculiarity also stems the
possibility for the surface stress-energy tensor to be NEC-violating,
in the sense that its timelike eigenvalue is not necessarily larger
than its spacelike eigenvalue, as should be the case in the more
restrictive field theory based situation. As a result, we have found
that the average values of the macroscopically relevant quantities are
unaltered by inclusion of dilatonic effects: the strings, from the
point of view of cosmology, are expected to form a Nambu-Goto-like
network.

However, the network of strings here produced would suffer from the
vorton excess problem~ \cite{vortons} because of the short-range
effects of the dilaton. As the stress-energy tensor eigenvalues depend
explicitly on the string worldsheet coordinates in a periodic way, it
is necessary, in order to form a string loop, that the wave number of
the dilatonic perturbation along the string be quantized. 

We also found that even though a short-distance cutoff must be
imposed, just like in the axionic situation~\cite{binetruy}, the
long-distance behavior of the dilaton can be adjusted so as to
suppress the logarithmic divergence. This means that a dilatonic
network of strings, apart from having an energy per unit length that
should be renormalized by inclusion of Planck scale effects, behaves
exactly as a usual network. Accordingly, all the results derived for
the latter~\cite{network} may be straightforwardly applied to the
former, in particular in regards of the predictions relative to the
loop production and the would-be ``vorton'' production.

More work needs to be done on such strings, as it is not clear in
particular whether there is any way to get rid of the ``current.'' If
no possibility can be found, that would mean that the vortons-like
configurations produced after the relevant phase transition would be
absolutely stable. As a result, one would have to conclude that cosmic
strings cannot be formed in the early universe if the underlying
theory of gravity is of the scalar-tensor type. Such models, stemming
from string theory, would therefore be incompatible with cosmic
strings; in view of the recently released CMB data~\cite{CMB}, this
might be presented as a useful constraint for the underlying
microscopic field theory.

\acknowledgments

V.~C.~de A. would like to thank FAPESP~--~BRASIL for financial
support.  M.~E.~X.~G. would like to acknowledge the kind
hospitality of the Institut d'Astrophysique de Paris where this work
has been completed. We would also like to thank J\'er\^ome Martin and
Raymond Schutz for enlightening comments and discussions.

\end{document}